\documentclass[aps,pra,reprint,floatfix,amsmath,amssymb,longbibliography]{revtex4-2}

\usepackage{braket}
\usepackage{graphicx}% Include figure files
\usepackage{dcolumn}% Align table columns on decimal point
\usepackage{bm}% bold math
\usepackage{hyperref}% add hypertext capabilities
\usepackage{enumitem}% itemize spacing
\usepackage{dsfont}% double-lined letters
\usepackage{csquotes}
\usepackage{float}
\usepackage{hyperref}
\usepackage{color}
\usepackage[dvipsnames]{xcolor}
\MakeOuterQuote{"}
\hypersetup{
    colorlinks=true,
    linkcolor={red!50!black},
    filecolor=blue,      
    urlcolor=blue,
    citecolor=blue,
}

\begin{document}
\title{Robustness of tripartite entangled states in passive PT-symmetric qubits}
\author{C. G. Feyisa\textsuperscript{1,2,3}}
\email{chimdessagashu@gmail.com}
\author{Cheng-Yu Liu\textsuperscript{1}}
\author{Muhammad S. Hasan\textsuperscript{1}}
\author{J. S. You\textsuperscript{4}}
%\email{jhihshihyou@gmail.com}
\author{Huan-Yu Ku\textsuperscript{4}}
\email{huan.yu@ntnu.edu.tw}
\author{H. H. Jen\textsuperscript{1,2,5}}
\email{sappyjen@gmail.com}
%\author{\textsuperscript{1,2}}
%\author{author4\textsuperscript{1}}
\affiliation{\textsuperscript{1}Institute of Atomic and Molecular Sciences, Academia Sinica, Taipei 10617, Taiwan}
\affiliation{\textsuperscript{2}Molecular Science and Technology Program, Taiwan International Graduate Program, Academia Sinica, Taiwan}
\affiliation{\textsuperscript{3}Department of Physics, National Central University, Taoyuan 320317, Taiwan}
\affiliation{\textsuperscript{4}Department of Physics, National Taiwan Normal University, Taipei 11677, Taiwan}
\affiliation{\textsuperscript{5}Physics Division, National Center for Theoretical Sciences, Taipei 10617, Taiwan}
\date{\today}% It is always \today, today,

\begin{abstract}
\noindent Non-Hermitian quantum systems have attracted significant interest in recent years due to the presence of unique spectral singularities known as exceptional points (EPs), where eigenvalues and eigenvectors coalesce. The drastic changes in these systems around their EPs have led to unique entanglement dynamics, which remained elusive until quite recently. In this work, we theoretically investigate the robustness of tripartite entanglement induced by EPs of the passive PT-symmetric non-Hermitian superconducting qubits, both in stand-alone configurations and hybrid setups with Hermitian qubits. In particular, we consider the qubits with both all-to-all and nearest-neighbour couplings under uniform and non-uniform coupling strengths. Our results reveal that non-Hermitian qubits with all-to-all coupling generate GHZ states, while those with nearest-neighbour interactions produce W states. These entangled states are resilient to non-uniform couplings and off-resonant driving fields. Moreover, the hybrid configurations combining Hermitian and non-Hermitian qubits suggest the importance of EPs for generating and maintaining genuine tripartite entanglement in our system. Additionally, driving the PT-symmetric qubits with a strong Rabi frequency can help sustain tripartite entanglement over time by countering losses, while strong inter-qubit coupling can benefit these entangled states in the low dissipation regime. These findings suggest that exploiting non-Hermitian systems and their associated EPs can create robust entangled states which are useful for both fundamental studies and quantum technologies.
\end{abstract}

\maketitle

\section{INTRODUCTION}
\label{introduction}
Non-Hermitian quantum systems have recently attracted much attention  due to the existence of spectral singularities called exceptional points (EPs) \cite{ar01, ar02, ar03, ar04, ar05, ar06, ar07, ar09, ar010, ar011, ar012, ar015, ar014, ar013, ar016}. These systems possess Liouvillian EPs when described by the Liouvillian super-operators or non-Hermitian EPs when governed by the non-Hermitian Hamiltonians \cite{ar017, ar018}. The difference between these EPs depends on whether the considered system dynamics includes or excludes effect of quantum jumps, respectively \cite{ar017,ar019}. An exclusive consideration of the non-Hermitian Hamiltonian is usually sufficient for open quantum systems upon post selection \cite{ar017}, and both the Liouvillian and non-Hermitian EPs are different from the diabolic points (DPs) where the eigenvalues coalesce but the corresponding eigenstates remain orthogonal \cite{berry:1984, arkhipov:2023}. In contrast, the loss of the conventional orthogonality condition at the EPs of the non-Hermitian systems leads to a real-complex spectral transition under both parity (${\cal P}$) and time reversal (${\cal T}$) transformations \cite{ozdemir:2019,ar011,ar012}. ${\cal P}{\cal T}$-symmetric systems \cite{bender:2024}, characterized by real spectra, have been implemented by simultaneously controlling gain and loss \cite{ar1, ar2, ar3} or just by controlling loss alone via Hamiltonian dilations \cite{ar4, ar5, ar6}. In the former approach, balancing gain and loss creates active ${\cal P}{\cal T}$-symmetric systems, while the latter, comprising only lossy components, results in passive ${\cal P}{\cal T}$-symmetric systems. 

Beyond their role in fundamental studies of physics in the classical and semi-classical regimes \cite{bender:2005, deffner:2015, Bender:2019}, 
${\cal P}{\cal T}$-symmetric systems \cite{bender:2024} have recently been shown to exhibit unique quantum entanglement dynamics, particularly near their EPs, due to the coalescence of multiple energy eigenstates \cite{ar7, ar10, ar14, ar8, ar9}. Entanglement of strongly coupled non-Hermitian qubits undergoes a spectral transition between the passive ${\cal P}{\cal T}$-symmetric and ${\cal P}{\cal T}$-broken phases \cite{ar8, ar9}. In the ${\cal P}{\cal T}$-symmetric phase, it exhibits oscillatory behaviour, whereas in the ${\cal P}{\cal T}$-broken phase, the entanglement remains stationary with finite strength. Moreover, externally driven and weakly coupled non-Hermitian qubits generate Bell state that is faster than their Hermitian counterparts \cite{ar7}. This result has been extended to multi-qubit systems \cite{ar10}, in which three qubits exhibit the same time-scaling as two qubits while four-qubit system further speeds up entanglement generation - showing a promising advantage of the higher-order EPs of many-body systems. 

Furthermore, involving an additional gain to the passive ${\cal P}{\cal T}$-symmetric qubits have shown to boost the success rate at the cost of large individual qubit driving \cite{ar14}. Such rapid entanglement generation and control are useful in practical settings, particularly for harnessing quantum effects within the coherence time \cite{ar11, ar12}. Moreover, fast entangled state generation ensures the reliability and accuracy of quantum computations and communications by reducing errors from qubit decoherence, which can otherwise degrade the performance of quantum information processing.

In this article, we investigate the robustness of EP-induced tripartite entanglement in non-Hermitian qubits and in hybrid setups consisting of both Hermitian and non-Hermitian qubits. We consider different coupling regimes, where the qubits are connected either through all-to-all or nearest-neighbour interactions, with both uniform and non-uniform coupling strengths. Our findings show that all-to-all coupled non-Hermitian qubits generate GHZ classes \cite{zeilinger:1992, bouwmeester:1999}, whereas those with nearest-neighbour coupling produce W classes \cite{eibl:2004,dur:2000}. These entangled states are developed over the same timescale despite the fundamental differences between GHZ and W states \cite{dur:2000}. Moreover, these states persist across a broad range of non-uniform inter-qubit coupling strengths and exhibit robustness against off-resonant driving effects. 

We further demonstrate that switching one of the non-Hermitian qubits to a Hermitian one results in biseparable entangled states due to the mismatched evolution times between Hermitian and non-Hermitian qubits. Additionally, we observe that increasing the Rabi frequency drives the system from the $\mathcal{PT}$-broken phase to the $\mathcal{PT}$-symmetric phase, where genuine tripartite entangled states can be generated rapidly and sustained for long time under strong driving fields. Finally, we show that fine-tuning the coupling constant from weak to strong regimes benefits non-Hermitian qubits only in the small dissipation regime, which suggests an entanglement growth in the Hermitian limit with negligible dissipation as the coupling strengths increase.

The manuscript is organized as follows: in Sec. \ref{model_method}, we introduce the Hamiltonian of the system and give the useful measure of entanglement. In Sec. \ref{all-to-all and nearest neighbour coupling} we discuss entanglement of non-Hermitian qubits with all-to-all and nearest-neighbour couplings. We then go into details of this entanglement against non-uniform couplings and off-resonant driving in Sec. \ref{non_uniform_coupling}, and entanglement generation in combined Hermitian and non-Hermitian qubits in Sec. \ref{combinations_H_and_NH_quibits}. We consider the behaviour of entanglement under strong coupling and driving in Sec. \ref{comparison_H_results}, and conclude in Sec. \ref{conclusions}.
\section{Hamiltonian of the system}
\label{model_method}
We consider a transmon circuit consisting of a capacitor and an inductor connected by a superconducting wire \cite{ar04,ar07,ar7}. Dynamics of the system is captured by the Hamiltonian written in the interaction picture as \cite{ar04,ar07,ar7}
\begin{eqnarray}
\hat{H}&=&\sum^{3}_{j=1}\bigg[(\Delta_j-\frac{i\gamma_j}{2})\hat{\sigma}_{j}\hat{\sigma}^{\dagger}_{j}+\Omega_{j}\hat{\sigma}^{x}_{j}\bigg]\nonumber\\&+&\sum^{3}_{j\neq k}\sum^{3}_{k=1}J_{jk}\big(\hat{\sigma}^{\dagger}_{j}\hat{\sigma}_{k}+\hat{\sigma}_{j}\hat{\sigma}^{\dagger}_{k}\big), \label{e1}
\end{eqnarray} 
where $\hat{\sigma}_j=|e\rangle_j \langle f|$ and $\hat{\sigma}_j^\dagger=|f\rangle_j \langle e|$ represent the ladder operators with $\lvert e \rangle_{j}$ and $\lvert f \rangle_{j}$ defining the first and second excited states of $j$th qubit,  $\hat{\sigma}_j^x=\hat{\sigma}_j^\dagger + \hat{\sigma}_j$ denotes the Pauli matrix, 
$J_{jk}$ indicates couplings between $j$th qubit and $k$th qubit, $\Omega_{j}$ designates an external driving frequency detuned by $\Delta_{j}$, and $\gamma_j$ is dissipation of level $|e\rangle_j$. The Hamiltonian obeys the passive ${\cal P}{\cal T}-$symmetric condition for $\Delta_j=0$ \cite{ar7,guo:2009}. 

The three-qubit system evolves under Eq. (\ref{e1}) according to 
\begin{eqnarray}
|\psi(t)\rangle=e^{-i\hat{H}t}|\psi(0)\rangle, \label{ee1}
\end{eqnarray}
in which $|\psi(0)\rangle$ is an arbitrary initial state. The state defined in Eq. (\ref{ee1}) is not normalized due to the intrinsic dissipation $\gamma_j$ introduced in Eq. (\ref{e1}). Therefore, we apply manual normalization, which amounts to performing subsequent post-selection in practice \cite{ar04, ar07}. The normalized state can also be expressed in a more convenient form as
\begin{eqnarray}
|\psi(t)\rangle=\sum^{8}_{k=1}a_{k}e^{-iE_kt}|\phi_k\rangle, \label{eee1}
\end{eqnarray}
where $a_k$ denotes a constant determined from the overlap of the initial state $|\psi(0)\rangle$ with the left eigenstates, $E_k$ represents the complex eigenvalues, and 
$|\phi_k\rangle$ are the normalized right eigenvectors in the biorthogonal bases \cite{ar7,bender:2024,bhosale:2021,brody:2013}. 

We can then expand Eq. (\ref{eee1}) in terms of the computational three-qubit bases as
\begin{eqnarray}
|\psi(t)\rangle=\sum^{8}_{k=1}\tilde{a}_k(t)|\alpha_k\rangle, \label{e3}
\end{eqnarray}
where $\tilde{a}_k(t)$ indicates a time dependent and complex probability amplitudes, and $|\alpha_k\rangle$ is the three-qubit computational bases arranged from $|eee\rangle$ for $k=1$ to $|fff\rangle$ for $k=8$.

%In general, our Hamiltonian can be non-symmetric depending on the values of $J_{jk}$. 
We next investigate the three-qubit entanglement using pairwise concurrences \cite{wootters:1998}, residual three tangle \cite{dur:2000, coffman:2000}, and von Neumann entropies of the reduced qubits \cite{islam:2015,ar10}. Vanishing pairwise concurrences suggest disappearance of bipartite entanglement among qubit pairs and emergence of multi-qubit GHZ states \cite{dur:2000}. On the other hand, pairwise concurrences remain finite for W states, reflecting that only the GHZ state has entanglement fully distributed among all three qubits with zero pairwise concurrences. Therefore, pairwise concurrences can be used to distinguish multiparty GHZ and W states. They can be calculated as \cite{wootters:1998, verstraete2001maximally}
\begin{eqnarray}
\mathcal{C}_{jk} = \text{max} \left\{0, \lambda_{1}-\lambda_{2}-\lambda_{3}-\lambda_{4} \right\},
\end{eqnarray}
where ${\lambda_i}$ denotes the eigenvalues of the Hermitian matrix $\sqrt{\sqrt{\hat\rho_{jk}(t)}\tilde{\hat\rho}_{ij}(t)\sqrt{\hat\rho_{jk}(t)}}$, in which $\tilde{\hat\rho}_{jk}(t)=(\hat\sigma_y\otimes\hat\sigma_y)\hat\rho^{\ast}_{jk}(t)(\hat\sigma_y\otimes\hat\sigma_y)$, $\hat\rho_{jk}$ indicates the joint density matrix of $j$th qubit and $k$th qubit, and $\hat\sigma_y$ is the Pauli $y$ matrix. 

The three-qubit GHZ and W states can also be identified by the residual three-tangle which is a genuine measure of tripartite entangled state defined as \cite{dur:2000, coffman:2000}.
\begin{eqnarray}
\tau=4|d_{1}-2d_{2}+4d_{3}|,\label{e4}
\end{eqnarray}
where $d_{1}=\left(\tilde{a}_{1}\tilde{a}_{8}\right)^{2} + \left(\tilde{a}_{2}\tilde{a}_{7}\right)^{2} + \left(\tilde{a}_{3}\tilde{a}_{6}\right)^{2} + \left(\tilde{a}_{4}\tilde{a}_{5}\right)^{2}$, $d_{2}=\tilde{a}_{1}\tilde{a}_{8}\left(\tilde{a}_{4}\tilde{a}_{5}+\tilde{a}_{3}\tilde{a}_{6}+\tilde{a}_{2}\tilde{a}_{7}\right)
+\tilde{a}_{3}\tilde{a}_{4}\tilde{a}_{5}\tilde{a}_{6} + \tilde{a}_{4}\tilde{a}_{5}\tilde{a}_{2}\tilde{a}_{7}+\tilde{a}_{2}\tilde{a}_{7}\tilde{a}_{3}\tilde{a}_{6}$, and 
$d_{3}=\tilde{a}_{1}\tilde{a}_{7}\tilde{a}_{4}\tilde{a}_{6} + \tilde{a}_{2}\tilde{a}_{8}\tilde{a}_{3}\tilde{a}_{5}.$ The residual three-tangle, i.e., Eq. (\ref{e4}), is symmetric under qubit permutations, vanishes for W and biseparable states, and becomes unity for GHZ states.

Furthermore, all three-qubit pure states given in Eq. (\ref{e3}) can, in fact, be locally transformed into a decomposition of five basis states \cite{acin2000generalized}, and they can further be classified as GHZ or W states using the entanglement (von Neumann) entropies of the reduced qubits. The entanglement entropy of $j$th qubit is given by \cite{islam:2015}
\begin{eqnarray}
\mathcal{S}_{j}=-{\rm Tr}[\hat\rho_{j}(t){\rm ln}\hat\rho_{j}(t)],
\end{eqnarray}
where $\hat{\rho}_{j}(t)$ indicates the reduced state of qubit $j$. Each qubit subsystem in GHZ states achieves the maximum entanglement entropies ${\cal S}_j={\rm ln}{2}$ while W states achieve ${\cal S}_j={\rm ln}{3}-(2/3){\rm ln}{2}\approx0.637$ \cite{li2022}. We next present our main results beginning with weakly coupled non-Hermitian qubits that possess eight-order EPs at $\Omega_j=\gamma_j/4$ for $J_{jk}=0$ \cite{ar10}. The regime $\Omega_j>\gamma_j/4$ corresponds to the passive ${\cal P}{\cal T}$-symmetric phase \cite{ar7,guo:2009} in our system.
\section{All-to-all and nearest-neighbour couplings} 
\label{all-to-all and nearest neighbour coupling} 
\begin{figure}[!t] 
%\includegraphics[width=1\textwidth]{00.pdf}\\ 
%\hspace{-3.5cm}(c)\hspace{5.5cm}(d)\hspace{6.5cm}(e)\\
\hspace{-2.75cm}(a)\hspace{4cm}\textcolor{white}{(b)}\\
\includegraphics[width=.485\textwidth]{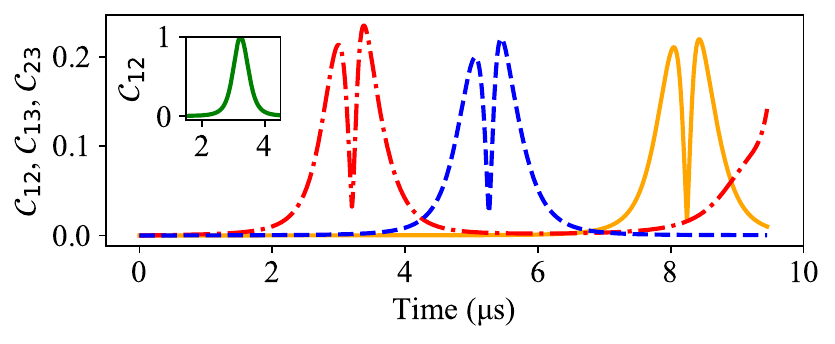}\\
\hspace{-2.75cm}(b)\hspace{4cm}\textcolor{white}{(b)}\\
\includegraphics[width=.485\textwidth]{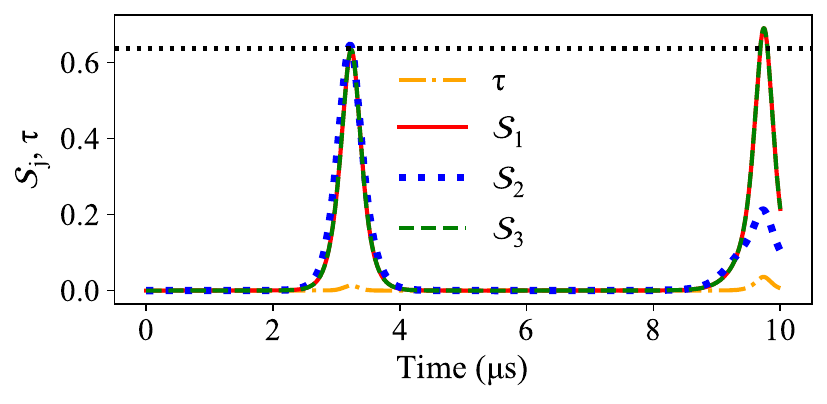}
%\hspace{-7.75cm}(b)\\
%\includegraphics[width=.5\textwidth]{co1.pdf}\\
%\includegraphics[width=.5\textwidth]{co2.pdf}\\
%\hspace{-2.75cm}(c)\hspace{4cm}\textcolor{white}{(b)}\\
%\includegraphics[width=.5\textwidth]{Nearest.pdf}\\
\caption{Dynamics of three non-Hermitian superconducting qubits from an initial state $|\psi(0)\rangle=2^{-3/2}(|f\rangle - i |e\rangle)^{\otimes 3}$. (a) Pairwise concurrences for $\Delta=0$, $\gamma=6$ rad/$\mu$s, $\{\Omega,J\}=\{1.512$ rad/$\mu s,10^{-5}$ rad/$\mu s\}$ (solid orange), $\{\Omega,J\}=\{1.529$ rad/$\mu s,10^{-4}$ rad/$\mu s\}$ (dotted blue), and $\{\Omega,J\}=\{1.576$ rad/$\mu s,10^{-3}$ rad/$\mu s\}$ (dash-dotted red). The inset shows bipartite entanglement (solid green) when the third qubit is disconnected from the composite system  as a comparison. % (b)  Entanglement entropies of Hermitian qubits corresponding to optimal regimes in (a). 
(b) Entanglement entropies [first qubit (solid red), second qubit (dotted blue) and third qubit (dashed green)], and residual three tangle (dash-dotted orange) of non-Hermitian qubits coupled with nearest neighbour coupling for $\{\Omega,J\}=\{1.576$ rad/$\mu s,10^{-3}$ rad/$\mu s\}$. Dotted-black line in (b) marks $\mathcal{S}_1=\mathcal{S}_2=\mathcal{S}_3=\text{ln}3-(2/3)\text{ln}2$ (see text) which is the case for maximally entangled three-qubit W state.} \label{f1}
\end{figure}

In this section, we first summarize entanglement of identical non-Hermitian qubits with all-to-all coupling as shown in Fig. \ref{f1}(a), and we then extend the discussion to the case with nearest-neighbour coupling in Fig. \ref{f1}(b). 

In Fig. \ref{f1}(a), we present pairwise concurrences $\mathcal{C}_{jk}$ of cyclically coupled non-Hermitian qubits for different optimal driving frequencies $\Omega$ and coupling strengths $J$. In these parameter regimes, pairwise concurrences $\mathcal{C}_{ij}$ immediately form narrow dips as a signature of highly entangled states produced at specific times. When one qubit decouples from the three-qubit system, the remaining non-Hermitian qubits become maximally entangled with each other. This is illustrated in the inset of Fig. \ref{f1}(a), where we show maximal entanglement between the first two qubits with $\mathcal{C}_{12} \approx 1$ at $t\approx3.23$ $\mu$s. On the other hand, the bipartite entanglement between any qubit pairs vanishes for cyclically coupled three qubits, and entanglement among the three-qubit system emerges with entanglement entropies ${\cal S}_{j}\approx\ln 2$ and ${\tau}\approx1$, which is the property of three-qubit GHZ state \cite{coffman:2000, dur:2000}. 

We next demonstrate three-qubit system with nearest-neighbour coupling as shown in Fig. \ref{f1}(b). % for $\{\Omega, J\}=\{1.576$ rad/$\mu$s, $10^{-3}$ rad/$\mu$s$\}$. 
In this setup, entanglement entropies ${\cal S}_{j}$ of the reduced qubits remain nearly identical for over $t=8$ $\mu$s that surpasses the evolution period $T_{\rm NHQ}=4\pi/\sqrt{16\Omega^2-\gamma^2}=6.5$ $\mu$s of the qubits. At $t=3.23$ $\mu$s, the qubits achieve entanglement entropies of $\mathcal{S}_1\approx\mathcal{S}_2\approx\mathcal{S}_3\approx\text{ln}$$3-(2/3)\text{ln}$$2$ which more likely characterizes a three-qubit W state and its classes. This can be further verified by a vanishing residual three tangle (see dash-dotted orange line and Refs. \cite{coffman:2000,dur:2000} as well). However, this property of W state disappears in the next entanglement revival phase around $T_{\rm NHQ}+3.23$ $\mu$s due to the build up of strong bipartite quantum correlation between the first and third qubits with $\mathcal{S}_1=\mathcal{S}_3 \approx \ln2$, while the middle qubit acting as a high-purity channel with smaller entanglement entropy.

The three-qubit entangled states discussed above are induced by the higher-order EP as the system transitions into the ${\cal P}{\cal T}$-symmetric regime. In proximity to this EP, the qubits exhibit increased sensitivity to their inter-qubit coupling, and this facilitates the emergence of tripartite entangled states through the redistribution of population and phase among the three qubits \cite{ar10}. Further approaching the higher-order EP from the ${\cal P}{\cal T}$-symmetric regime requires smaller coupling strengths and longer timescales to generate two- and three-qubit entangled states \cite{ar7, ar10}. This can be attributed to the fact that the response of the qubits to their inter-qubit coupling weakens for smaller coupling strengths, and the qubits slowly evolve near the higher-order EPs. 

Although entanglement generation experiences delays near the EPs (solid orange line in Fig. \ref{f1}(a)), it remains significantly faster compared to the timescales of Hermitian qubits. The timescale for the latter is inversely proportional to the inter-qubit coupling strengths \cite{majer:2007} and can be related to the timescale of the all-to-all coupled non-Hermitian qubits as $t^{\rm Opt}_{\rm NHQs}\propto J_{jk}\times t^{\rm Opt}_{\rm HQs}$, where $t^{\rm Opt}_{\rm NHQs}$ and $t^{\rm Opt}_{\rm HQs}$ represent the optimal entanglement generation times for the non-Hermitian and Hermitian qubits, respectively. Furthermore, our numerical results indicate that the timescale $t^{\rm Opt}_{\rm NHQs}$ is approximately $50\%$ of the complete evolution period $T_{\rm NHQ}$ when the non-Hermitian qubits are initialized in coherent superpositions $|\psi(0)\rangle=2^{-3/2}(|f\rangle - i |e\rangle)^{\otimes 3}$ and approaches $T_{\rm NHQ}$ when initialized in the state $|\psi(0)\rangle=|fff\rangle$ \cite{ar10}. 

Moreover, our system can achieve both bipartite and tripartite entangled states within the same timescale (see Figs. \ref{f1}(a) and \ref{f1}(b), including the inset), regardless of the differences in EP orders between two and three non-Hermitian qubits. However, this situation may change for multiple-qubit configurations. For example, higher-order EPs of four qubits can further reduce timescale of entanglement generation below that of two and three non-Hermitian qubits \cite{ar10}. 
\section{Robustness to non-uniform couplings and off-resonant drivings} 
\label{non_uniform_coupling} 
\begin{figure}[!t] 
%\hspace{-1.55cm}(a)\hspace{5cm}\textcolor{white}{(b)}\\
%\hspace{-0.62cm}(c)\hspace{17cm}\textcolor{white}{(b)}\\
\hspace{-2.0cm}(a)\hspace{4cm}\textcolor{white}{(b)}\\
\includegraphics[width=0.485\textwidth]{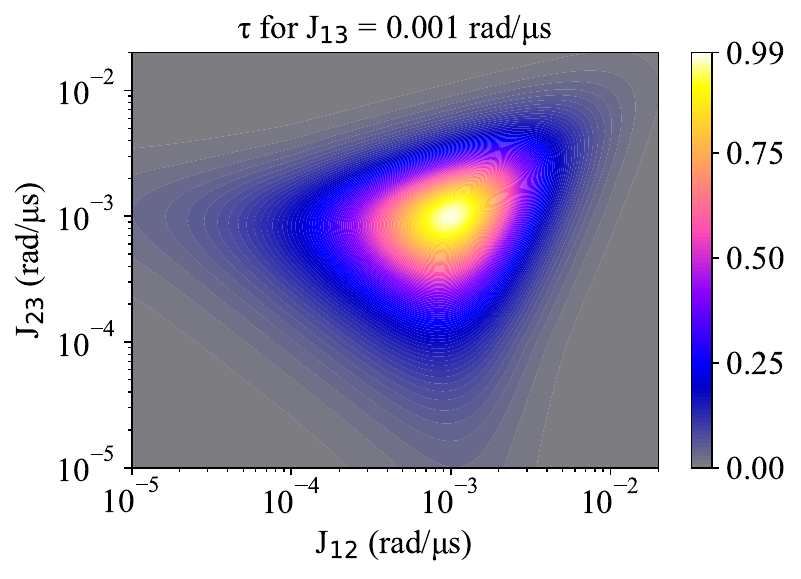}\\ 
\hspace{-2.55cm}(b)\hspace{4cm}\textcolor{white}{(b)}\\
\includegraphics[width=0.485\textwidth]{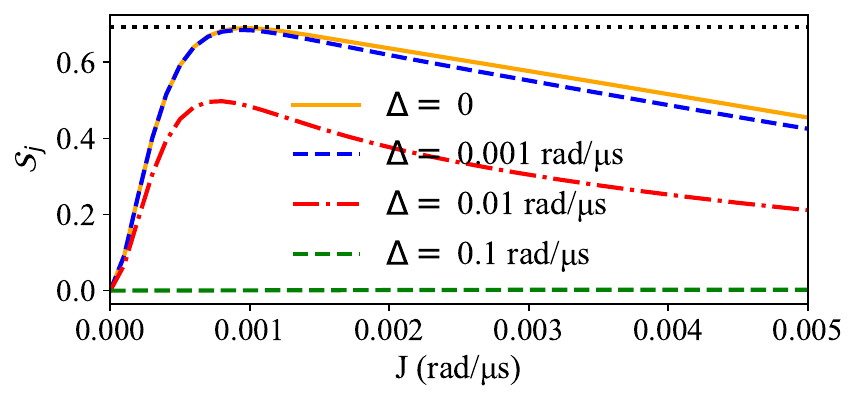}\\
\hspace{-2.55cm}(c)\hspace{4cm}\textcolor{white}{(b)}\\
\includegraphics[width=.485\textwidth]{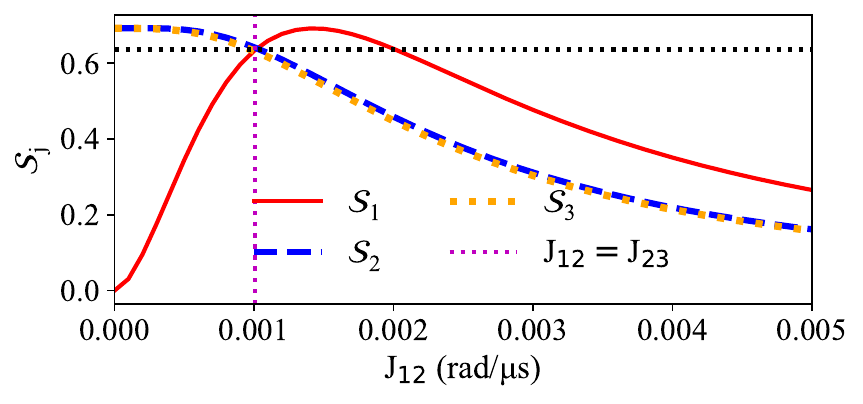}
%\hspace{-4.5cm}(f)\hspace{5.25cm}(g)\hspace{5.25cm}(h)\\
%\includegraphics[width=1\textwidth]{com.pdf}  
\caption{Effect of non-uniform coupling and off-resonant driving on the entanglement of non-Hermitian qubits. (a) Residual three-tangle of non-uniformly coupled non-Hermitian
 qubits with all-to-all configuration for $\{\Omega, J_{13}, t\} = \{1.576 \text{ rad}/\mu\text{s}, 10^{-3} \text{ rad}/\mu\text{s}, 3.23$  $\mu\text{s}\}$. (b) Robustness of optimal entanglement (bright region in (a)) under off-resonant driving for $\Delta = 10^{-3}$ (dashed blue), $10^{-2}$ (dash-dotted red), and $10^{-1} \text{ rad}/\mu\text{s}$ (dotted green). The resonant case $\Delta = 0$ (solid orange) is shown for comparison. (c) Entanglement with non-uniform nearest-neighbour coupling, $\{\Omega, J_{23}, J_{13}, t\} = \{1.576 \text{ rad}/\mu\text{s}, 10^{-3} \text{ rad}/\mu\text{s}, 0, 3.23$ $\mu\text{s}\}$. Dotted dark lines indicate upper bounds (b) $\mathcal{S}_j=\ln 2$ for a GHZ state and (c) $\mathcal{S}_j=\ln 3-(2/3)\ln 2$ for a W state (see text), while the dotted magenta line in (c) represents uniform nearest-neighbour coupling ($J_{12}=J_{23}$) for optimal W state generation. The initial state is the same as in Fig.~\ref{f1}, and rest of the parameters are set to $\Delta = 0$, $\Omega = 1.576 \text{ rad}/\mu\text{s}$, $\gamma = 6 \text{ rad}/\mu\text{s}$ unless otherwise specified.}
 \label{f2}
\end{figure}

Next we discuss the robustness of entanglement in non-Hermitian qubits against non-uniform couplings and off-resonant drivings. We demonstrate the residual three-tangle for all-to-all coupled non-Hermitian qubits in Fig. \ref{f2}(a) at $t \approx 3.23$ $\mu$s, and we find that finite entangled states persist over a broad range of non-uniform inter-qubit couplings but vanish with larger coupling variations. As shown earlier in Sec. \ref{all-to-all and nearest neighbour coupling}, this setup produces a GHZ state with $\tau \approx 1$ when $J_{jk} \approx 10^{-3}$ rad/$\mu$s. Under this condition, we present robustness of the state to off-resonant driving in Fig. \ref{f2}(b), where we show that entanglement remains consistent with the resonant case as long as $\Delta \lesssim J^{\rm Opt}$ with an optimal coupling $J^{\rm Opt}=J^{\rm Opt}_{jk}=10^{-3}$ rad/$\mu$s. However, large detuning frequencies significantly exceeding an optimal inter-qubit coupling $J^{\rm Opt}$ hampers the entanglement generation.

Furthermore, in Fig.~\ref{f2}(c), we illustrate entanglement entropies of the reduced qubits at $t=3.23$ $\mu$s for the nearest-coupling ($J_{13}=0$) and for finite $J_{23}$ as we vary $J_{12}$. When $J_{12}=0$, this three-qubit configuration only supports bipartite entanglement between the last two qubits, as the first qubit is decoupled from the composite system (see Sec. \ref{all-to-all and nearest neighbour coupling} also). Coupling the first qubit to the second qubit and increasing $J_{12}$ slightly reduces this bipartite entanglement, while the purity of the first qubit decreases rapidly, suggesting possible three-qubit entanglement generation. Interestingly, at $J_{12}=J_{23}=10^{-3}$ rad/$\mu$s, the entanglement entropies of the reduced qubits converge to nearly equal values $\mathcal{S}_j\approx\text{ln}$$3-(2/3)\text{ln}$$2$ and the residual three tangle almost vanishes there (see Fig. \ref{f1}(b)). This indicates that optimal W state generation also requires uniform nearest-neighbour coupling, and the robustness of entanglement to non-uniform coupling is evident from Fig. \ref{f2}(c) where the decreased entanglement entropies of the second and third qubits are compensated by an increased entanglement entropy of the first qubit over a wide range of $J_{12}$ especially near $\mathcal{S}_j\approx\text{ln}$$3-(2/3)\text{ln}$$2$. Meanwhile, further increasing $J_{12}$ does not benefit entanglement generation.
\section{Hybrid setups of Hermitian and non-Hermitian qubits} 
\label{combinations_H_and_NH_quibits} 
\begin{figure}[!t] 
\centering
\hspace{-2.75cm}(a)\hspace{3.5cm}{(b)}\\
\includegraphics[width=0.485\textwidth]{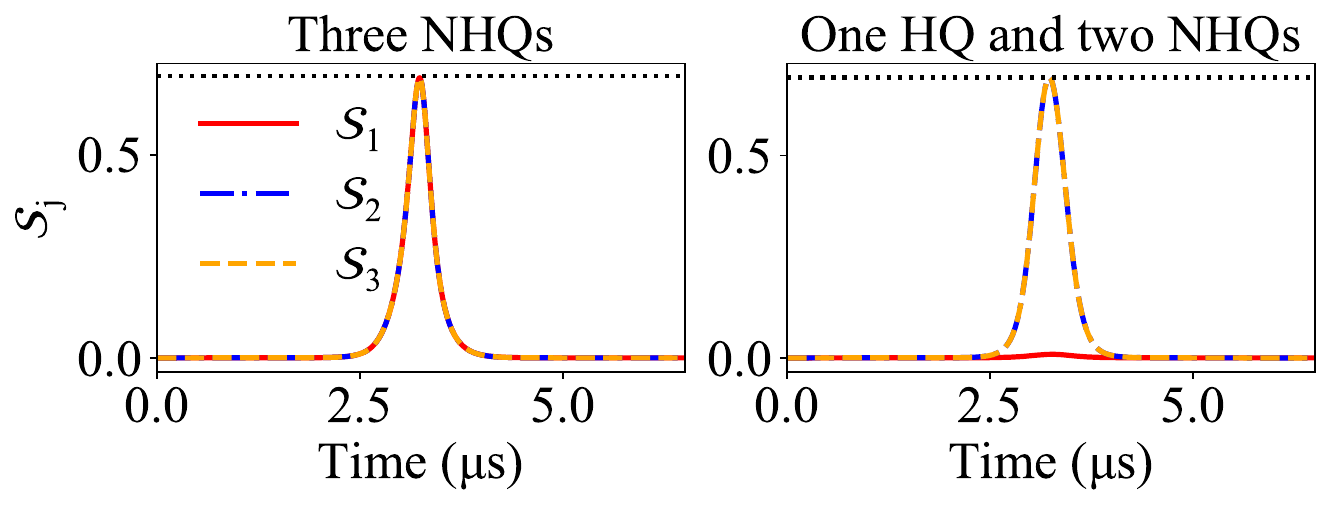}\\
\hspace{-2.75cm}(c)\hspace{3.5cm}{(d)}\\
\includegraphics[width=0.485\textwidth]{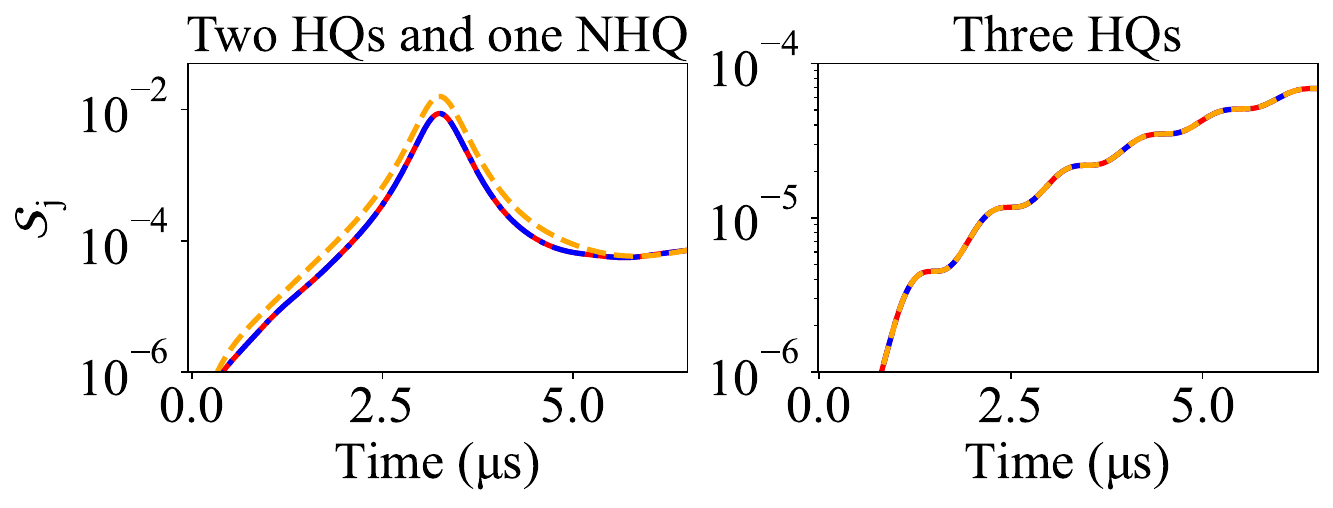}
%\includegraphics[width=0.5\textwidth]{comm.pdf}
%\hspace{-2.55cm}(c)\hspace{4cm}\textcolor{white}{(b)}\\
%\includegraphics[width=0.5\textwidth]{commm.pdf}
\caption{Dynamics of quantum entanglement in the hybrid setups of non-Hermitian (NHQ) and Hermitian (HQ) qubits. (a-d) Entanglement entropies (a) for all non-Hermitian qubits, (b) for one Hermitian (solid red) and two non-Hermitian qubits (dash-dotted blue and dashed orange), (c) for two Hermitian qubits (solid red and dash-dotted blue) and one non-Hermitian qubits (dashed orange), and (d) for three identical Hermitian qubits. The initial state is the same as in Fig. \ref{f1}, and the rest parameters are $\Delta=0$, $J_{jk}=10^{-3}$ rad/$\mu$s, $\Omega=1.576$ rad/$\mu$s and $\gamma=6$ rad/$\mu$s.}
\label{f3}
\end{figure}
Coupling one Hermitian qubit with a non-Hermitian one leads to a rich landscape of EPs that facilitate maximal bipartite entanglement generation \cite{kumar:2022} and enable realization of a quantum Hatano-Nelson model with asymmetric hopping \cite{hatano:1996}. We now extend the discussion of tripartite entanglement generation in hybrid configurations of Hermitian and non-Hermitian qubits. These setups can help distinguish the advantage of introducing non-Hermitian effect in our setup and its associated EPs on tripartite entanglement generation. 

We first demonstrate in Fig. \ref{f3}(a) that the three-non-Hermitian qubits generate genuine and strong tripartite entangled states around $t=3.23$ $\mu$s (see Ref. \cite{ar10} for the details). In Fig. \ref{f3}(b), we show entanglement entropies of one Hermitian and two non-Hermitian qubits, where the non-Hermitian qubits are maximally entangled and the Hermitian qubit remains in separable state (${\cal S}_1\approx0$). Without coupling, this qubit-arrangement possesses two symmetrical fourth-order EPs at $\Omega=1.5$ rad/$\mu$s, and one of the EPs induces rapid entanglement build up between the non-Hermitian qubits when they are driven by $\Omega=1.576$ rad/$\mu$s and uniformly coupled with $J_{jk}=10^{-3}$ rad$/\mu$s, while the other EP is redundant and plays no role in the entanglement generation. % as if it emerges from the expanded Hilbert space due to the presence of the Hermitian qubit. 

Moreover, the evolution times of the Hermitian and the two non-Hermitian qubits are incompatible due to the introduced non-Hermitian effects. Therefore, switching one of the qubits to be Hermitian will not generate genuine tripartite entanglement entanglement but only bipartite entangled state. This approach can be extended to rapidly generate multi-qubit biseparable entangled states of GHZ classes, which can keep entanglement even if some qubits are detached similar to multiparty W states upon particle loss. Such biseparable entangled states can also be prepared by carefully controlling inter-qubit coupling, e.g., by setting $J_{23}=J_{13}=0$ in three qubit system while $J_{12}$ is finite.  

Moreover, we show in Fig. \ref{f3}(c) that a setup of two Hermitian and one non-Hermitian qubits further reduces the strength of the entanglement, but we can still notice the signature of non-Hermitian qubit at $t=3.23$ $\mu$s and entanglement entropies can be larger than that of all three Hermitian qubits as shown in Fig. \ref{f3}(d). Such combinations of Hermitian and non-Hermitian qubits highlight the importance of higher-order EPs of the multi-qubit system to achieve and maintain multipartite entangled states.
\section{Strong coupling and driving effects}
\label{comparison_H_results}
\begin{figure}
\begin{center}
\hspace{-3.25cm}(a)\hspace{4.25cm}{(b)}\\
\includegraphics[width=0.485\textwidth]{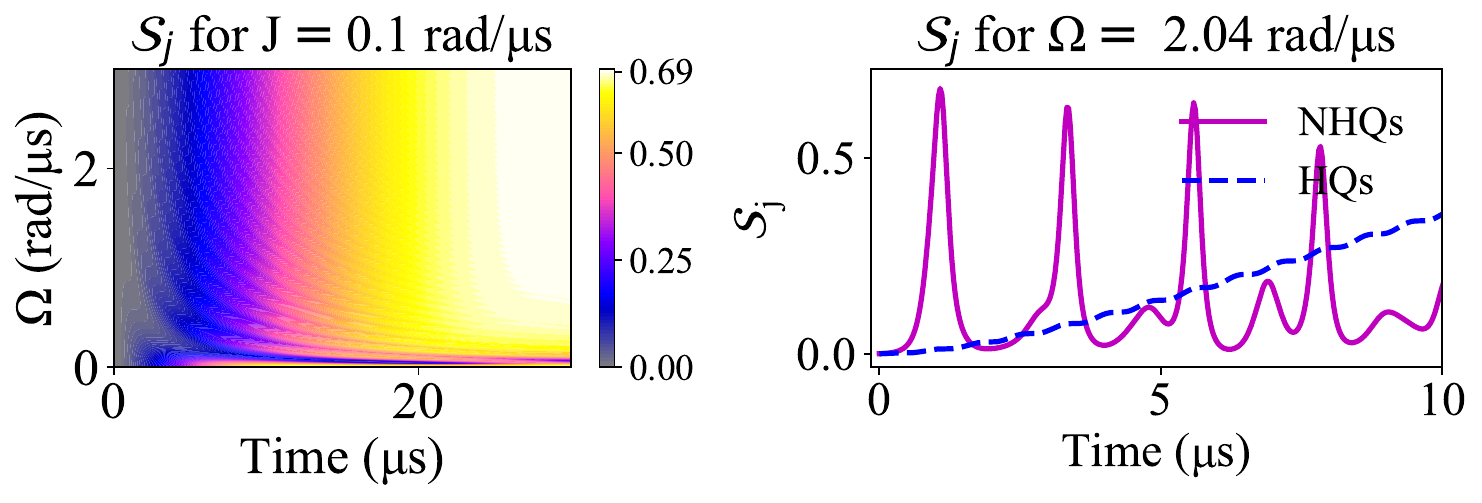}\\
\hspace{-2.9cm}(c)\hspace{3.75cm}{(d)}\\
\includegraphics[width=0.485\textwidth]{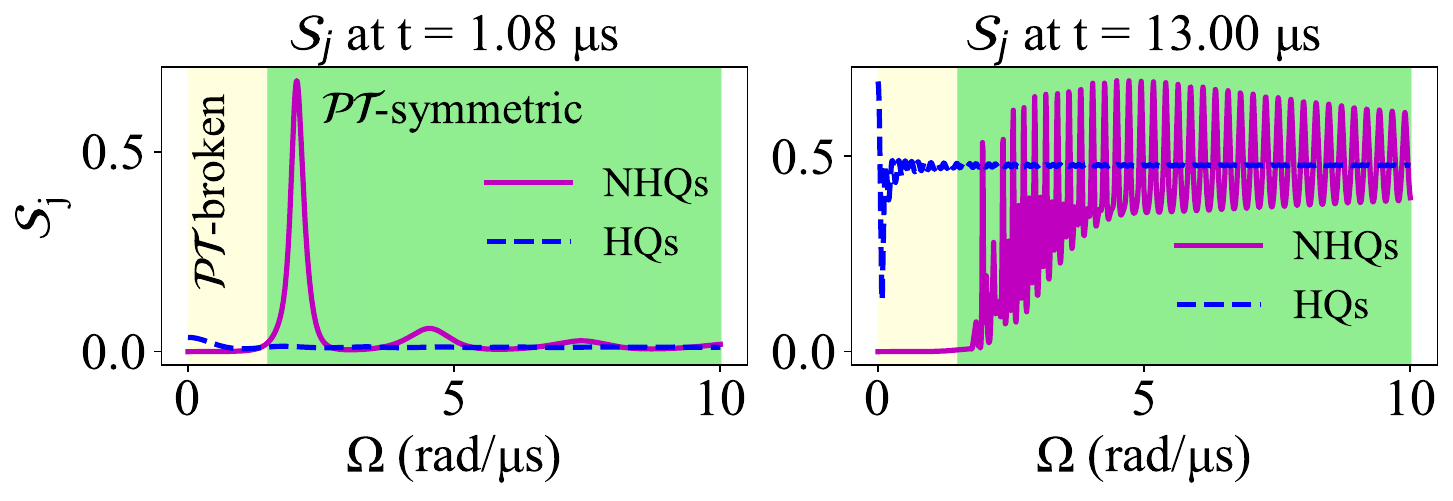}
\caption{Effect of strong coupling and strong driving on entanglement of the non-Hermitian qubits from the same initial state in Fig. \ref{f1}. (a) Entanglement entropy of reduced Hermitian qubits with parameters $J=0.1$ rad$/\mu$s, $\Delta=0$, and $\gamma=0$. (b) Comparison of entanglement entropies for reduced Hermitian qubits (dashed blue, $\gamma=0$) and non-Hermitian qubits (solid magenta, $\gamma=6$ rad$/\mu$s), under conditions $J_{jk}=0.1$ rad$/\mu$s, $\Delta=0$, $\Omega=2.04$ rad$/\mu$s, and $\gamma=6$ rad$/\mu$s. (c, d) Entanglement entropy of reduced qubits as a function of Rabi frequency: (c) at an early stage of the dynamics ($t=1.08$ $\mu$s) and (d) at a later stage ($t=13$ $\mu$s). The system undergoes ${\cal P}{\cal T}$ protected spectral transition from broken phase (light yellow) to passive symmetric phase (light green) by tuning the Rabi frequency as shown (c) and (d).}\label{f5}
\end{center} 
\end{figure} 
In the previous section, we have shown that three identical coupled non-Hermitian qubits are necessary to speed up tripartite entanglement generation. We next make detail analysis of this entanglement by considering strong driving regime or strong coupling regime. We increase the coupling constant to $J=0.1$ rad/$\mu$s for which the timescale of entanglement generation in the non-Hermitian qubits is still better than that of the Hermitian qubits. 

We first consider entanglement of all-to-all coupled Hermitian qubits in Fig. \ref{f5}(a) as a reference. These qubits produce GHZ states nearly at $13$ $\mu$s even without driving them. Exposing the individual qubits to the driving field can help generate a three-qubit GHZ state and its classes that remain stable for long time. In Fig. \ref{f5}(b), we compare entanglement of Hermitian and non-Hermitian qubits under an optimal driving amplitude $\Omega=2.04$ rad/$\mu$s associated to the coupling $J_{jk}=0.1$ rad$/\mu$s. Non-Hermitian qubits generate entangled states that periodically oscillate with time, and stronger entangled state appears earlier around $t=1.08$ $\mu$s. This timescale is about $20$ times faster than the rate of entanglement generation in the driven Hermitian qubits. Furthermore, Fig. \ref{f5}(c) illustrates that further increasing the Rabi frequency drives the non-Hermitian qubits from the ${\cal P}{\cal T}$-broken (light yellow) to the ${\cal P}{\cal T}$-symmetric phase (light green), causing rapid entanglement growth, while Hermitian qubits show negligible entanglement.

The entanglement entropy of qubits at a later stage of dynamics is shown in Fig. \ref{f5}(d) with an example at $t=13$ $\mu$s. Hermitian qubits become entangled even without driving and form stable entangled states with finite strength when driven. In contrast, non-Hermitian qubits fail to generate entanglement in the ${\cal P}{\cal T}$-broken regime due to significant losses over time that require strong driving to counteract. This indicates that the effects of non-Hermiticity, including EPs, stem from the competition between driven and dissipative processes in our system. For strong driving, these entangled states oscillate rapidly and follow the Hermitian result for $\Omega\geq\gamma$. We have also checked that only specific coupling strengths can yield entanglement in non-Hermitian qubits, as increasing the coupling constant alone does not overcome the loss, unlike driving. In contrast, Hermitian qubits exhibit increased entanglement with higher coupling strengths.
\section{Conclusions}
\label{conclusions}
In conclusion, we have discussed the robustness of EP-induced tripartite entanglement in passive ${\cal P}{\cal T}$-symmetric non-Hermitian qubits and hybrid systems made up of Hermitian and non-Hermitian qubits. We considered all-to-all and nearest-neighbour interactions with uniform and non-uniform coupling strengths and examined strong driving effects where non-Hermitian qubits undergo spectral transitions protected by ${\cal P}{\cal T}$-symmetry. 

We find that all-to-all coupled non-Hermitian qubits rapidly generate GHZ states which are characterized by narrow dips in pairwise concurrences, upper bounds of both residual three-tangle and entanglement entropies. In contrast, nearest-neighbour coupled qubits produce W states on the same timescale. Furthermore, these entangled states are robust against off-resonant driving effects and can persist across a broad range of non-uniform coupling strengths, with optimal conditions being achieved under uniform coupling. In hybrid setups consisting Hermitian and non-Hermitian qubits, the introduced non-Hermiticity causes incompatible evolution times and prevents the creation of genuine tripartite entangled states, but biseparable entangled states can still be induced in the hybrid systems. This indicates that non-Hermiticity and its associated EPs are crucial to generate, speed up, and maintain multipartite entangled states. 

We have also shown that driving the qubits with a strong Rabi frequency can sustain these entangled states in the ${\cal P}{\cal T}$-symmetric phase by counteracting losses, while strong inter-qubit coupling enhances the entanglement in the low dissipation regime. This suggests that the effects of non-Hermiticity, including EPs, arise from the interplay between driven and dissipative processes in our system.

While our study focuses on passive ${\cal P}{\cal T}$-symmetric qubits, future research could consider the effects of quantum jumps \cite{ar017, ar018} or incorporate an additional gain to our system \cite{ar01, ar14, ar012}, particularly to enhance success rates of entanglement generation. Another promising direction would be the realization of the quantum Hatano-Nelson model in hybrid setups of Hermitian and non-Hermitian qubits \cite{hatano:1996, kumar:2022}. Our model could also be implemented in superconducting circuits \cite{ar07,ar04,roy:2020} and microwave quantum state router \cite{zhou:2023}. Furthermore, our system can be naturally extended to investigate multipartite entangled states and their phase transition induced by skin effect \cite{ar06}.

\begin{acknowledgments}
H. H. J. acknowledges support from the National Science and Technology Council (NSTC), Taiwan, under Grant No. NSTC-112-2119-M-001-007 and
Grant No. NSTC-112-2112-M-001-079-MY3, from Academia Sinica under Grant No. AS-CDA-113-M04, Taiwan, and from TG 1.2 of NCTS at NTU, Taiwan. H.- Y. K. is supported by the Ministry of Science and Technology, Taiwan, (with grant number MOST112-2112-M-003- 020-MY3), and Higher Education Sprout Project of National Taiwan Normal University (NTNU).
\end{acknowledgments}

\bibliography{apssamp}

\end{document}